\documentclass[twocolumn,twocolappendix]{aastex631}
\usepackage{natbib}

\usepackage{amsmath}
\usepackage{algorithm}
\usepackage{amsfonts}
\usepackage{mathrsfs}
\usepackage{amssymb}
\usepackage{color}
\usepackage{graphicx} 
\let\tablenum\relax
\usepackage{siunitx}
\usepackage{comment}

\newcommand{\cha}{\textit{Chandra}}

\def \UXClumpy {{\tt \UXClumpy}}

\begin{document}
\title{Fast X-ray Variability from the Coronae of Supermassive Black Holes}

\author[0000-0002-7791-3671]{Xiurui Zhao}
\affiliation{Cahill Center for Astrophysics, California Institute of Technology, 1216 East California Boulevard, Pasadena, CA 91125, USA}
\affiliation{Department of Astronomy, University of Illinois at Urbana-Champaign, Urbana, IL 61801, USA}

\author[0000-0001-8822-8031]{Luca Comisso}
\affiliation{Department of Astronomy and Columbia Astrophysics Laboratory, Columbia University, 538 West 120th Street, New York, NY 10027, USA}

\author[0000-0002-2203-7889]{Stefano Marchesi}
\affiliation{Dipartimento di Fisica e Astronomia (DIFA), Università di Bologna, via Gobetti 93/2, 40129 Bologna, Italy}
\affiliation{Department of Physics and Astronomy, Clemson University, Kinard Lab of Physics, Clemson, SC 29634, USA}
\affiliation{INAF, Osservatorio di Astrofisica e Scienza dello Spazio di Bologna, via P. Gobetti 93/3, 40129 Bologna, Italy}

\author[0000-0002-6584-1703]{Marco Ajello}
\affiliation{Department of Physics and Astronomy, Clemson University, Kinard Lab of Physics, Clemson, SC 29634, USA}

\author[0000-0002-0273-218X]{Elias Kammoun}
\affiliation{Cahill Center for Astrophysics, California Institute of Technology, 1216 East California Boulevard, Pasadena, CA 91125, USA}

\author[0000-0003-1659-7035]{Yue Shen}
\affiliation{Department of Astronomy, University of Illinois at Urbana-Champaign, Urbana, IL 61801, USA}


\author[0000-0003-4202-1232]{Qiaoya Wu}
\affiliation{Department of Astronomy, University of Illinois at Urbana-Champaign, Urbana, IL 61801, USA}

\begin{abstract}
We present the first systematic study of short-timescale X-ray variability in radio-quiet active galactic nuclei (AGN), utilizing archival \textit{Chandra} observations of approximately 3,000 broad-line AGN selected from the SDSS and DESI spectroscopic surveys. We identify 14 AGN exhibiting rapid {(on timescales of tens of kiloseconds)} X-ray flux variations by factors of two or more that are statistically significant {($p\le6\times10^{-4}$)}, indicative of fast coronal variability. By converting minimum variability timescales to light-crossing times, we place upper limits on the sizes of the variable coronal regions, finding typical scales of $\lesssim10^{-4}$~pc. The coronal variable region size upper limits of {\bf an} AGN in our sample are found to be much smaller than the typical coronal sizes inferred from microlensing, suggesting that its corona is composed of localized, transient structures rather than smooth, homogeneous plasmas. Such efficient magnetic energy dissipation in compact volumes is consistent with expectations for magnetically dominated coronae and is supported by recent general relativistic magnetohydrodynamic simulations. Future high-throughput X-ray observatories will enable the detection of even faster coronal variability, providing direct constraints on the physical mechanisms driving plasma energization and flux fluctuation near supermassive black holes. Our results suggest that luminous AGN hosting massive black holes are prime targets for probing the small-scale structure and dynamics of AGN coronae.  
\end{abstract}


\section{Introduction}\label{sec:intro}
Active galactic nuclei (AGN) are among the most luminous persistent sources in the Universe across the electromagnetic spectrum, powered by accretion onto supermassive black holes (SMBHs) at their centers. X-ray emission in the $\sim$0.1--100~keV range contributes approximately 5--30\% of the total bolometric luminosity of AGN, depending on their intrinsic luminosity and accretion rate. Notably, AGN are the dominant contributors to the cosmic X-ray background \citep[CXB,][]{Giacconi1962}, underscoring the importance of understanding their high-energy emission processes.

The majority of AGN X-ray emission, particularly above $\sim$2~keV, is thought to originate from a hot ($T_c \sim 10^8$--$10^{10}$~K) corona of ionized plasma that resides in the innermost region. In this region, ultraviolet and optical photons from the disk are up-scattered via inverse Compton processes into the X-ray regime. The corona is extremely compact, so its size cannot be directly resolved in current X-ray imaging. Indirect constraints, primarily from microlensing observations \citep[e.g.,][]{Morgan2008,Chartas2009,Dai2010,MacLeod2015,Chartas2016}, suggest that the radius of the corona spans a few tens of gravitational radii ($r_{\rm g} = GM_{\rm BH}/c^2$, where $M_{\rm BH}$ is the SMBH mass, $G$ is the gravitational constant, and $c$ is the speed of light).

X-ray emission of AGN is known to be variable across a wide range of timescales, from minutes to years, reflecting the dynamic nature of the corona \citep[see][for recent reviews]{Laha2025,Paolillo2025}. On timescales of days to years, AGN exhibit stochastic (red-noise) variability, with larger amplitudes at longer timescales \citep[e.g.,][]{Uttley2005,Gonzlez2012}. This long-timescale behavior is often attributed to inward-propagating fluctuations in the accretion flow. However, significant short-term variability on timescales of kiloseconds (ks) has not been systematically investigated, and its physical origin remains poorly understood.

Recent years have witnessed significant progress in the theoretical understanding and numerical modeling of AGN coronae. General relativistic magnetohydrodynamic (GRMHD) simulations of accretion flows indicate that magnetic fields threading the accreting material can cause the inner regions to transition at small radii into a hot, magnetically dominated corona near the black hole \citep{Liska2022}. Dissipation of these magnetic fields through reconnection \citep{Rowan2017} and turbulence \citep{Comisso2019} can energize the plasma, sustaining the high temperatures inferred from X-ray observations. However, these ideas remain largely speculative, and observational constraints are limited. In particular, characterizing coronal variability, especially on short timescales, is crucial for testing theoretical scenarios, as variability encodes information about the energy release, characteristic size scales, and dynamical processes within the corona. Systematic studies of rapid X-ray variability can thus provide direct constraints on the physical mechanisms governing coronal plasma energization and flaring near SMBHs.

{The origin of variability, particularly in the X-ray band, remains one of the fundamental open questions in AGN physics. Various mechanisms have been proposed to explain the observed rapid X-ray variability, including magnetic reconnection, accretion disk instabilities, failed jets, non-stationary coronae, and eclipses by intervening material. Each of these mechanisms may operate under different conditions, but their relative importance across the AGN population is still uncertain \citet[see,][for a recent overview of AGN variability from both observational and theoretical perspectives]{Paolillo2025}. In this work, we aim to derive new observational constraints on short-timescale variability, providing a foundation for future efforts to relate variability properties to the physical state of the coronal plasma and the accretion flow.}

In this work, we present the first systematic study of short-timescale (on timescales of tens of ks) X-ray variability in radio-quiet AGN, utilizing archival \textit{Chandra} observations of nearly 3,000 broad-line AGN selected from the SDSS and DESI spectroscopic surveys. This paper is presented as follows: in Section~\ref{sec:selection}, we present the sample selection criteria when systematically studying the fast coronal variability. In Section~\ref{sec:DA}, we present the analysis of the X-ray and optical data of the sources in our sample, where we derived their X-ray lightcurves, black hole masses, and X-ray luminosities. We also present the properties of the detected variability of these sources. In Section~\ref{sec:discussion}, we discuss the discovered fast coronal variability, the observed correlation between black hole mass and variable region size, and the gravitational and relativistic effects on the variability timescales. 
All uncertainties quoted in this paper are at the 1$\sigma$ (68\%) confidence level unless otherwise stated. Standard cosmological parameters are adopted as follows: $H_0$ = 70 km s$^{-1}$ Mpc$^{-1}$, $\Omega_M$ = 0.3, and $\Omega_\Lambda$ = 0.7.

\section{Source Selection}\label{sec:selection}
The AGN analyzed in this study were selected from the largest optical spectroscopic surveys of AGN to date: the Sloan Digital Sky Survey (SDSS) Data Release 16 AGN catalog \citep{Wu2022} and the Dark Energy Spectroscopic Instrument (DESI) Data Release 1 AGN catalog \citep{DESI2025}. Together, these catalogs contain over 2.5 million unique broad-line AGN, with more than 750,000 from SDSS and 2.1 million from DESI. We focus exclusively on broad-line (Type 1) AGN, as their emission line widths are required to estimate black hole masses and, hence, the gravitational radii of their central SMBHs. 

To identify AGN with significant short-term X-ray variability, we cross-matched these AGN catalogs with the \cha\ Source Catalog (CSC) version 2.1.1 \citep{Evans2024}, which contains uniformly processed data from over two decades of \cha\ observations. The CSC provides calibrated source properties, including positions, fluxes, and variability metrics. As this study targets variability on timescales of a few to tens of ks, we restricted our analysis to intra-observation variability, using only individual observations. Variability probabilities in the CSC are calculated using three statistical methods: the Gregory--Loredo algorithm, the Kolmogorov--Smirnov test, and Kuiper's test.\footnote{\url{https://cxc.cfa.harvard.edu/csc/columns/variability.html}}

We matched the CSC to SDSS and DESI AGN using a 1\arcsec\ radius to ensure reliable associations \citep[e.g.,][]{Marchesi2016}. To ensure adequate photon statistics, only AGN with more than 100 net counts in a single \cha\ observation were considered. This yielded 2,950 unique AGN observed in {4,875} total exposures. We then selected only sources with variability probabilities exceeding 99\% in the full band (0.5--7~keV) across all three statistical methods as listed in CSC.

To isolate coronal variability, we excluded radio-loud AGN with radio-loudness $RL>10$ using the NRAO VLA Sky Survey \citep[NVSS,][]{Condon1998} data, as their X-ray emission may be dominated by jet-related processes. We also removed sources whose background light curves showed similar variability trends to the source. We further remove targets located near galaxy clusters to mitigate potential lensing-related distortions. In total, 39 AGN with 69 significantly variable observations were initially selected. We performed visual inspection of the background light curves of these 69 observations and excluded potential contamination from solar flares, radiation belt crossings, and cosmic rays. Therefore, the final sample comprises 19 distinct AGN across 22 observations.

\section{Data Analysis and Results}\label{sec:DA}
\subsection{Data Analysis}
While the \cha\ Source Catalog (CSC) provides pre-generated light curves for each X-ray detection, these are constructed using the Gregory--Loredo algorithm with adaptive, weighted binning optimized for variability detection. However, such binning is not well-suited for time-domain analyses aimed at quantifying variability timescales, as in this work. To perform a consistent and temporally resolved analysis, we re-extracted light curves for all 19 significantly variable AGN using the standard CIAO pipeline\footnote{\url{https://cxc.cfa.harvard.edu/ciao/threads/lightcurve/}}. {The light curves were initially generated using a 100~s bin.} Source regions were defined as circular apertures with a 10\arcsec\ radius, while background spectra were extracted from nearby 30\arcsec\ circular regions on the same detector chip, free from contamination by other X-ray sources.

To search for variability over different timescales and enhance the signal-to-noise ratio (S/N), we regrouped each observation using time intervals of 3, 6, 9, 12, and 15 ks. {The regrouping starts from any point between 100~s and the chosen bin size (e.g., for a 3~ks bin, the first interval can begin at any time between 100 s and 3~ks, followed by subsequent 3~ks intervals), allowing sufficient flexibility in time sampling. We consider the variability significant if the difference in count rates between any two time bins within the observation exceeds a 4$\sigma$ threshold. The upper and lower uncertainty of the count rate is determined using Equations (9) and (12) in \citet{Gehrels1986} with $S$ = 1 for 1\,$\sigma$ confidence level.} 17 AGN present more than 4$\sigma$ variabilities in 20 of their archival observations. We define the minimum variability timescale ($t_{\rm min}$) as the shortest interval with flux variability by a at least factor of two identified among the significant ($>$4$\sigma$) variations in the light curves, as usually applied in previous works \citep[e.g.,][]{Boller1997,Aharonian2007,Sbarrato2011}. Four sources did not exhibit flux variations exceeding a factor of two during their observations. Consequently, 14 AGN across 16 observations showed significant ($>4\,\sigma$) flux variability exceeding this threshold. In particular, J1238+6213 displayed such variability in three separate archival observations.

\begingroup
\renewcommand*{\arraystretch}{1.2}
\begin{table*}
\begin{center}
\caption{Summary of source properties of the selected variable AGN and their \cha\ observations. The black hole mass of the sources measured in this work is labeled with stars * ({see details in Section~\ref{sec:BH}}) and the black hole masses of the remaining sources are adopted from \citet{Wu2022}. CR is the average net count rate of the entire \cha\ observation. {We report the median value and the 16th--84th percentile ranges of LCT.} $R_{\rm CVR}$ is {the 3~$\sigma$} upper limit of the radius of the coronal variable region in units of both $r\rm _g$ {\bf (derived from the simulations)} and pc.}\label{Table:CS_UL}
  \begin{tabular}{ccccccccccccc}
       \hline
       \hline
    Source&$z$&log(M$\rm_{BH}$)&{log($r\rm _g$)}&ObsID&{CR}&$p$&$sig$&$ratio$&$t_{\rm min}$&{LCT}&{\bf $R_{\rm CVR}$}&$R_{\rm CVR}$\\ 
    &&M$_{\odot}$&{light-sec}&&cts/ks&10$^{-3}$&$\sigma$&&ks&$r\rm _g$/$c$&$r\rm _g$&10$^{-4}$~pc\\
       \hline
        \hline
J0811+0231&2.546&9.4$\pm$0.5&4.1$\pm$0.5&16687&2.9&0.5&4.1&3.4&12$\pm${4.9}&{0.24$_{-0.17}^{+0.56}$}&{\bf $<$9.0}&{$<$0.73}\\
J1130+3640&1.353&9.1$\pm$0.5&3.8$\pm$0.5&06945&3.1&$<$0.1&4.5&4.2&12$\pm${4.9}&{0.84$_{-0.60}^{+2.0}$}&{\bf $<$32}&{$<$1.1}\\
J0832+5243&1.580&8.4$\pm$0.5*&3.1$\pm$0.5&07684&2.2&0.2&4.3&3.6&12$\pm${4.9}&{3.3$_{-2.4}^{+7.8}$}&{\bf $<$130}&{$<$1.0}\\
J1349+2601&1.226&8.7$\pm$0.5&3.4$\pm$0.5&17609&5.9&0.6&4.6&2.5&{36$\pm$4.9}&{7.0$_{-4.8}^{+15}$}&{\bf $<$230}&{$<$2.2}\\
J1235+1227&0.809&8.0$\pm$0.5&2.7$\pm$0.5&13985&2.9&0.1&4.1&3.9&9$\pm${3.6}&{9.0$_{-6.4}^{+21}$}&{\bf $<$340}&{$<$1.1}\\
J1000+0204&1.234&8.2$\pm$0.5&2.9$\pm$0.5&08012&16&0.3&4.7&2.4&{20$\pm$2.0}&{11$_{-7.7}^{+22}$}&{\bf $<$340}&{$<$1.1}\\
J1128+5837&0.781&7.8$\pm$0.5&2.5$\pm$0.5&01641&6.3&0.4&4.4&2.3&12$\pm${4.9}&{20$_{-14}^{+48}$}&{\bf $<$770}&{$<$1.3}\\
J1532+3017&0.214&8.0$\pm$0.5&2.6$\pm$0.5&14009&4.4&$<$0.1&4.5&2.7&12$\pm${4.9}&{21$_{-15}^{+49}$}&{\bf $<$790}&{$<$2.2}\\
J1238+6213&0.441&8.0$\pm$0.5&2.7$\pm$0.5&03390&5.3&0.6&4.6&2.1&{15$\pm$6.1}&{22$_{-16}^{+51}$}&{\bf $<$820}&{$<$2.2}\\
-&-&-&-&03388&6.3&0.5&4.6&2.6&{27$\pm$3.7}&{41$_{-28}^{+90}$}&{\bf $<$1300}&{$<$2.6}\\
-&-&-&-&03391&6.3&0.2&5.2&2.3&{120$\pm$6.1}&{190$_{-130}^{+390}$}&{\bf $<$5900}&{$<$9.3}\\
J1602+4303&0.072&7.8$\pm$0.5*&2.5$\pm$0.5&03460&39&$<$0.1&4.6&2.1&9$\pm${1.2}&{30$_{-21}^{+63}$}&{\bf $<$950}&{$<$1.1}\\
J1215+4713&0.497&8.0$\pm$0.5&2.7$\pm$0.5&04738&2.1&$<$0.1&4.6&6.9&24$\pm${4.9}&{33$_{-23}^{+73}$}&{\bf $<$1100}&{$<$2.5}\\
J1604+4303&0.506&7.6$\pm$0.5&2.3$\pm$0.5&06932&3.2&0.2&4.1&3.9&12$\pm${4.9}&{40$_{-29}^{+95}$}&{\bf $<$1500}&{$<$1.7}\\
J1426+3506&0.217&7.7$\pm$0.5*&2.4$\pm$0.5&16321&29&0.3&4.4&2.0&15$\pm${1.2}&{45$_{-30}^{+95}$}&{\bf $<$1400}&{$<$1.5}\\
J1415--0030&0.135&7.0$\pm$0.5*&1.7$\pm$0.5&12906&88&$<$0.1&9.4&2.3&21$\pm${1.2}&{390$_{-270}^{+810}$}&{\bf $<$12000}&{$<$2.1}\\
       \hline
\end{tabular}
\end{center}
\end{table*}
\endgroup

{To further assess the significance of the detected variability using our adopted criteria, namely, that the count rate varies by more than a factor of two and exceeds the $4\sigma$ threshold between any two time bins within an observation, and to determine the minimum variability timescale ($t_{\rm min}$) while ensuring that the number of false detections due to statistical fluctuations remains low, we performed a Monte Carlo simulation. Specifically, we generated 10,000 simulated light curves for each observation of the 14 selected sources under the null hypothesis of no intrinsic variability. The light curves were produced with a 100~s bin, consistent with the analysis of the real observations. For each bin, the number of source and background counts was drawn from Poisson distributions based on the average source and background count rates measured in the real observations. The simulated light curves were then grouped into intervals of 3, 6, 9, 12, and 15 ks, and variability tests were conducted following the same procedure described above.

We computed the $p$-value as the fraction of simulated light curves that exhibit a count-rate variation greater than or equal to that measured in the real observation and that also exceed the observed significance threshold between any two time bins. To limit the expected number of false detections, we required $p \lesssim 6\times10^{-4}$, corresponding to upto three spurious variable sources among the 4,875 sources in our sample. Because each observation can exhibit various combinations of count-rate variability amplitude (${\it ratio}$) and significance (${\it sig}$) across different $t_{\rm min}$ values, we evaluated all combinations to identify the minimum $t_{\rm min}$ that satisfies the $p \lesssim 6\times10^{-4}$ criterion. For each AGN and observation, the selected minimum variability timescale, along with the corresponding ${\it sig}$ and $p$-value, is listed in Table~\ref{Table:CS_UL}. Considering the derived $p$-values for each source and the total number of sources in our sample, we estimate that approximately one out of the 16 detected variable observations may be due to statistical fluctuations.} The corresponding light curves are shown in Fig.~\ref{fig:LC1}--\ref{fig:LC3} in the Appendix. It is worth noting that even faster variability ($<$3~ks), involving flux variability by a at least factor of two, may be present among the targets. However, such signals are unlikely to reach high statistical significance (e.g., $>4\sigma$) given the limited quality of the current data set (see further discussion in Section\ref{sec:MBH_RCVR}).

{It is worth noting that the expected number of spurious variable sources may be decreased due to the presence of radio-loud AGN among the 4,875 sources (which were not excluded when cross-matching the CSC and SDSS/DESI samples). In addition, the intrinsically stochastic variability of AGN on timescales of tens to hundreds of ks may increase the expected number of spurious detections. A detailed investigation of these effects, which are likely to be minimal, is left to future studies.}

\subsection{Results}\label{sec:results}
We found 14 AGN exhibiting significant short-term X-ray variability at the $\ge$4$\sigma$ level in their 16 \cha\ archival observations. The minimum variability timescale is listed in Table~\ref{Table:CS_UL}. {The physical light-crossing time (LCT) of the coronal variability region is expressed in units $r_{\rm g}/c$, which is calculated by dividing the physical size of the coronal variability region (in unit of $r_{\rm g}$) converted from the rest-frame minimum variability timescale ($t_{\rm min}$) by the speed of light ($c$), i.e., LCT = $c\,t_{\rm min}(1+z)^{-1}(r_{\rm g})^{-1}c^{-1}$.The uncertainty in the LCT is estimated using Monte Carlo simulations, where 10 million realizations of $t_{\rm min}$ and $r_{\rm g}$ are generated to compute the corresponding LCT values. For each source, we report the median value and the 16th--84th percentile range of the resulting LCT distribution in Table~\ref{Table:CS_UL}. It is worth noting that the large uncertainty in the LCT primarily arises from the uncertainty in the black hole mass.}

In reality, photons emitted during the variability are expected to undergo scattering and diffusion within the coronal plasma, resulting in effective propagation speeds slower than the speed of light. Thus, by assuming free-streaming photon propagation at light speed, the observed variability timescales provide upper limits on the physical size of the flaring region. We refer to this constraint as the coronal variable region (CVR) upper limit, where the radius of the CVR in unit of parsec (pc) is $R_{\rm CVR}<c\,t_{\rm min}(1+z)^{-1}$, assuming the coronal variable region to be a spherical blob. {The upper limit of $R_{\rm CVR}$ in unit of $r_{\rm g}$ is calculated by the 84th percentile value of LCT distribution (LCT$_{84}$) multiplied by $c$. The 3~$\sigma$ upper limits of $R_{\rm CVR}$ in units of both $r_{\rm g}$ and pc are reported in Table~\ref{Table:CS_UL}. } The gravitational and relativistic effects might affect the estimation of the CVR region size, which will be discussed in Section~\ref{sec:GR}.

We find that the majority of R$_{\rm CVR}$ upper limit estimates are comparable to, or much smaller than, the typical coronal sizes inferred from microlensing studies, which are on the order of $\sim$10--100~$r_{\rm g}$ (Fig.~\ref{fig:LCT_relation} and Table~\ref{Table:CS_micro} in Appendix). {\bf Notably, one source exhibits $R_{\rm CVR}$ 3~$\sigma$ upper limit values $<$10~$r_{\rm g}$.} This implies extremely compact flaring regions, significantly smaller than the global coronal extent, and is suggestive of localized, high-energy activity in these sources. The compact flaring regions also support the clumpy coronae scenarios \citep[e.g.,][]{Haardt1991,Jiang2019}.

{We also examined the hardness ratio (HR) of each source as a function of time, where the hard (H) and soft (S) bands correspond to the 2--10 keV and 0.5--2 keV count rates, respectively, and defined as ${\rm HR} = (H - S) / (H + S)$. For most sources, the HR does not exhibit significant ($<2\sigma$) variations. An exception is J1415--0030, which shows a softer-when-brighter trend at $\sim3\sigma$ significance, consistent with either an increase in accretion rate or the clearing of an obscuring structure along the line of sight. Overall, there is no compelling evidence that the observed variability in most sources is driven by changes in absorption.}

{The energetic X-ray variability in short-timescale found in AGN suggests a rarely-probed class of X-ray variable phenomena, which is further discussed in Section~\ref{sec:rare}.}

\section{Discussion}\label{sec:discussion}
\subsection{Fast Coronal Variability}
The very short timescale (tens of ks) variability uncovered in this work requires the presence of compact, high-energy-density regions within the corona that are capable of rapid energy dissipation. Flux variations exceeding a factor of two relative to quiescent levels indicate that the variable regions contribute substantially to the total coronal X-ray luminosity. {The 3~$\sigma$ upper limits on the radii of the variable coronal regions for the lower black hole mass (log(M$_{\rm BH}$)$<9$) sources in our sample are broadly consistent with the global coronal extents inferred from microlensing studies. In contrast, the very short $t_{\rm min}$ values and extremely compact variable regions observed in the high black hole mass systems, when compared to the microlensing-based global coronal sizes, strongly strongly disfavor homogeneous coronal models.} Instead, they favor scenarios where the corona is highly inhomogeneous, where magnetically dominated turbulence and magnetic reconnection could drive rapid energy release in localized regions. Continued monitoring of such rapid variability is essential for advancing our understanding of the physical processes operating in the immediate vicinity of SMBHs and for constraining the geometry of AGN coronae. Notably, Table.~\ref{Table:CS_UL} shows that luminous AGN hosting massive black holes, such as J0811+0231, represent especially promising laboratories for probing fast coronal variability in terms of light-crossing times. 

A large fraction of the sources in the sample with small $t_{\rm min}$ have low net count rate in their \cha\ observations. Therefore, it is highly possible that we will be able to observe much shorter $t_{\rm min}$ for the same events as observed in this work with future, more sensitive X-ray telescopes, thus placing tighter constraints on the size of the coronal variable region.

Recent GRMHD simulations \citep[e.g.,][]{Chashkina2021,Porth2021,Ripperda2022} predict the formation of such transient, magnetically dominated structures near the black hole, but the short durations and compact scales of the observed variability place stringent constraints on the underlying dissipation physics. Their energy output and timescales must be accounted for by the energy conversion mechanism (e.g., magnetic reconnection) occurring in the magnetized coronal plasma. However, the inferred extreme compactness of some variabilities ($<2\,r_g$) may require additional mechanisms, such as relativistic beaming (see Section~\ref{sec:GR}), to fully explain their observed properties.

The short timescale variability is assumed to be mainly caused by the reconnection of the local magnetic loops in the corona. While the thermal or magneto-rotational instabilities (MRI) in the innermost disk (at a few $r_{\rm g}$) might create localized bursts of UV emission on an hourly timescale.

\begin{figure}
\centering
\includegraphics[width=.48\textwidth]{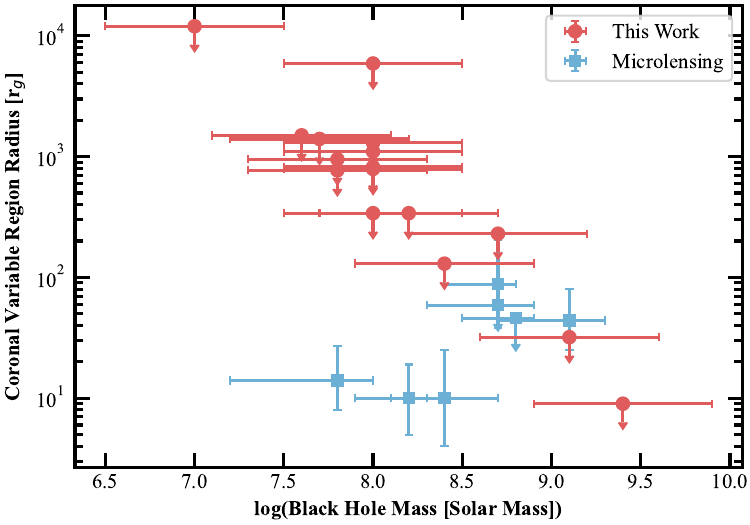}
\includegraphics[width=.48\textwidth]{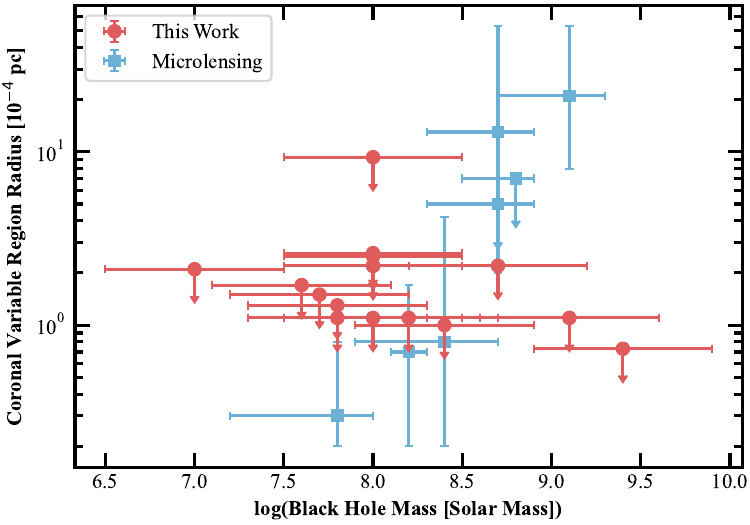}
\caption{{\bf 3~$\sigma$ upper limits of coronal variable region radius ($R_{\rm CVR}$) in units of r$_{g}$ (upper)} and pc (lower) measured in this work (red) and coronal size measured by microlensing method \citep[blue, adopted from][]{Chartas2016} as a function of the source black hole mass.}\label{fig:LCT_relation}
\end{figure}
\subsection{$M_{BH}$ vs $R_{\rm CVR}$}\label{sec:MBH_RCVR}

We find that sources hosting more massive black holes yield tighter constraints on the size of the coronal variable region (Fig.\ref{fig:LCT_relation}). This apparent correlation may arise from the combined effects of source flux and telescope sensitivity. In principle, for low-mass black holes ($\sim$10$^{7}$~M$\odot$), detecting $R_{\rm CVR}\lesssim$1~$r_{\rm g}$ would require observing variability on timescales as short as $\sim$10--100~s, owing to their much smaller $r_{\rm g}$. Such rapid variations are challenging to capture with the current data quality. For instance, we identify flux variability by at least a factor of two on $t_{\rm min}$ = 300~s, corresponding to $R_{\rm CVR}\lesssim$17~$r_{\rm g}$, in the lowest-mass source in our sample (also with the highest average flux), J1415--0030, but only at $\sim$2.3$\sigma$ significance. Therefore, observations with more sensitive telescopes should enable substantially tighter upper limits on $R_{\rm CVR}$ in AGN with low-mass black holes.

{The black hole masses of high-$z$ ($z\gtrsim$2) AGN in \citet{Wu2022} were estimated using the C IV $\lambda$1549 emission line. However, the C IV-based method is known to possibly overestimate the black hole mass relative to the H$\alpha$-based measurements, owing to the contribution of non-virial outflows and calibration uncertainties. The overestimation can reach up to a factor of $\sim$5--10, depending on the blueshift of the line \citep[e.g.,][]{Coatman2017}. To be conservative, we assume that the C IV-derived mass of J0811 is overestimated by a factor of 3 (which is highly unlikely as the blueshift of its C IV line is only about 600~km~s$^{-1}$, suggesting that its black hole mass is consistent with or even underestimated by $\sim$30\% compared with those expected from the H$\alpha$ method). Its true black hole mass would then be $\log(M_{\rm BH}/M_\odot) = 8.9 \pm 0.5$, corresponding to a light-crossing time of LCT = $1.2^{+2.8}_{-0.84}\,r{\rm g}$ and a $R_{\rm CVR}<$\,27$r_{\rm g}$. These values remain below the typical coronal sizes inferred from microlensing studies at comparable black hole masses. This conclusion is further reinforced by the coronal variable region size expressed directly in pc unit (Fig.~\ref{fig:LCT_relation}), which are entirely independent of black hole mass estimates. } 

{\bf Future better measured black hole masses of the AGN in the sample from e.g., reverberation mapping technique \citep{Shen2024} or interferometric observations of the broad-line region \citep{GRAVITY2017} would place more stringent constraints on the upper limits of the coronal variable region sizes.}

\subsection{Gravitational and Relativistic Effects} \label{sec:GR}
AGN coronae are known to be highly compact, spanning only a few tens of $r_{\rm g}$, and are located in the immediate vicinity of the central SMBH \citep[see][and references therein]{Reis2013}. This proximity places the coronal region deep within the strong gravity regime, where relativistic effects such as gravitational time dilation, light bending, and Doppler boosting become significant and must be taken into account when interpreting coronal properties.

\begin{itemize}
\item The observed duration of a variation, $t_{\rm obs}$, is longer than the intrinsic (local) duration, $\tau$, due to gravitational time dilation: $t_{\rm obs}$ = $\tau$/(1--2/$r$)$^{0.5}$, where $r$ is the radial distance of the coronal variability region from the SMBH in $r_{\rm g}$. For $r = 15\,r_{\rm g}$, the median coronal size inferred from microlensing, the time dilation effect is $\sim$7\% (i.e., $t_{\rm obs} \approx 1.07\,\tau$). For variabilities occurring closer to the black hole, e.g., at $r = 6\,r_{\rm g}$ (the innermost stable circular orbit for a non-spinning black hole, the time dilation increases to $\sim$23\% ($t_{\rm obs} \approx 1.23\,\tau$).

\vspace{0.1cm}

\item The intense gravitational field near the SMBH bends the trajectories of photons emitted from the corona. A significant fraction of this emission is redirected toward the accretion disk, enhancing the reflected component while suppressing the directly observed coronal continuum. For spatially extended coronae, photons emitted from regions closer to the SMBH travel longer curved paths than those emitted farther out, introducing GR-induced time delays. These delays can smear out fast variability features, particularly if the variability regions extend across a range of radii.

\vspace{0.1cm}

\item When the coronal flare possesses bulk motion toward the observer \citep[e.g.,][]{Ewing2025}, relativistic Doppler effects amplify the observed flux {and contract the timescale}. The enhancement factor for flux is given by $\mathcal{D}^{3+\alpha}$, where $\alpha$ is the spectral index of the emission and $\mathcal{D}$ is the Doppler factor: $\mathcal{D}$ = 1/$\gamma$/(1--$\beta$cos$\theta$) with $\beta = v/c$, $\gamma = 1/\sqrt{1 - \beta^2}$, and $\theta$ the angle between the velocity vector and the observer’s line of sight. {The time contraction is $t_{\rm obs} = \tau/\mathcal{D}$.} For example, a flare moving outward at $v = 0.1\,c$ along the line-of-sight with $\alpha = 1$ yields a flux enhancement of $\sim$49\%, and a time contraction of $\sim$10\% ($t_{\rm obs} \approx 0.90\,\tau$). A faster outflow, such as $v = 0.5\,c$ along the line-of-sight, leads to a more dramatic effect: a flux amplification by a factor of $\sim$9 and $t_{\rm obs} \approx 0.58\,\tau$.

\end{itemize}

Therefore, general relativistic effects introduce complex distortions to the intrinsic timescales and amplitudes of coronal variations occurring near the SMBH. In general, the net impact on the observed timescale is modest, as gravitational time dilation and Doppler time contraction partially counterbalance each other. However, Doppler boosting can significantly amplify the observed variability amplitude, particularly if the flaring region has substantial bulk motion along the line of sight. These effects must be carefully considered when interpreting the temporal and energetic properties of X-ray variabilities in AGN.

\section{Conclusion}\label{sec:conclusion}
We have presented the first systematic study of short-timescale X-ray variability in AGN across a broad redshift range and spanning black hole masses from $10^7$ to $10^{10}\,M_\odot$. Out of approximately 3,000 broad-line (type 1) AGN with \textit{Chandra} observations, 14 sources exhibit significant flux variability on timescales of tens of ks. The small number of detections suggests either that such energetic variabilities are intrinsically uncommon or that many short-timescale variabilities remain undetected due to the current instrumental sensitivity limits. 

Future high-throughput X-ray observatories, such as {\it AXIS} \citep{Reynolds2023}, {\it eXTP} \citep{Zhang2025}, and {\it NewAthena} \citep{Cruise2025}, will offer significantly improved sensitivity to lower-amplitude and faster variability, enabling a more complete census of coronal activity and allowing direct comparisons with theoretical models and simulations of magnetic reconnection and plasma turbulence in AGN coronae.

We also demonstrate that the minimum observed variability timescales provide meaningful upper limits on the size of individual coronal variable regions. {\bf In one source, these upper limits (in units of both $r_{\rm g}$ and pc) fall below the typical coronal extents inferred from microlensing, favoring the scenario that coronae of such source is }composed of localized, transient structures rather than smooth, homogeneous plasmas.

\section {Acknowledgements}
{The authors thank the anonymous referee for their insightful comments on the manuscript.} XZ appreciates the helpful discussion with Yan-Fei Jiang.

XZ acknowledges support from NASA grant 80NSSC24K1031. LC acknowledges support from NASA grant 80NSSC24K1230. MA acknowledges funding from NASA under contracts 80NSSC22K1579, 80NSSC24K1403, and 80NSSC24K1745.

{This paper employs a list of Chandra datasets, obtained by the Chandra X-ray Observatory, contained in~\dataset[DOI: 10.25574/cdc.501]{https://doi.org/10.25574/cdc.501}."}

This research has made use of data obtained from the \cha\ Source Catalog, provided by the \cha\ X-ray Center (CXC). This work makes use of the data from SDSS. Funding for the Sloan Digital Sky Survey has been provided by the Alfred P. Sloan Foundation, the U.S. Department of Energy Office of Science, and the Participating Institutions. DESI construction and operations are managed by the Lawrence Berkeley National Laboratory. This research is supported by the U.S. Department of Energy, Office of Science, Office of High-Energy Physics, under Contract No. DE–AC02–05CH11231, and by the National Energy Research Scientific Computing Center, a DOE Office of Science User Facility under the same contract. Additional support for DESI is provided by the U.S. National Science Foundation, Division of Astronomical Sciences, under Contract No. AST-0950945 to the NSF’s National Optical-Infrared Astronomy Research Laboratory; the Science and Technology Facilities Council of the United Kingdom; the Gordon and Betty Moore Foundation; the Heising-Simons Foundation; the French Alternative Energies and Atomic Energy Commission (CEA); the National Council of Science and Technology of Mexico (CONACYT); the Ministry of Science and Innovation of Spain, and by the DESI Member Institutions. The DESI collaboration is honored to be permitted to conduct astronomical research on Iolkam Du’ag (Kitt Peak), a mountain with particular significance to the Tohono O’odham Nation.

\appendix
\renewcommand{\thesection}{APPENDIX~\Alph{section}}
\section{Black Hole Mass Measurement}\label{sec:BH}
Black hole masses for SDSS AGN were obtained from \citet{Wu2022}, where they were estimated using the single-epoch spectroscopic method. This technique relies on empirical correlations between the broad emission-line width and the continuum luminosity, assuming virial motion of the line-emitting gas. One source in our sample, J0832+5243, lacks a reported black hole mass due to the absence of a 135~nm continuum luminosity measurement, as the spectrum does not cover this wavelength. For this object, we extrapolated the ultraviolet continuum and estimated the luminosity at 135~nm, enabling us to compute its black hole mass following the same methodology used in \citet{Wu2022}. The resulting value is listed in Table~\ref{Table:CS_UL}.

For AGN drawn from the DESI catalog, which do not have published black hole mass estimates, we derived virial masses using the same approach and calibration as in \citet{Wu2022}. These sources are identified accordingly in Table~\ref{Table:CS_UL}.

While the measurement uncertainties for single-epoch virial masses are typically small ($<$0.1 dex), the method is subject to systematic uncertainties of up to $\sim$0.4 dex \citep{Shen2013}. To account for both statistical and systematic effects conservatively, we adopt a uniform uncertainty of 0.5 dex for all black hole mass estimates in our sample.

\begingroup
\renewcommand*{\arraystretch}{1.2}
\begin{table}
\begin{center}
\small
\caption{Information of the selected variable AGN. The coordinates are measured in SDSS or DESI. Luminosities are reported in logarithm.}\label{Table:source_property}
  \begin{tabular}{cccccc}
       \hline
       \hline
    Source&RA&DEC&L$\rm_{2-10}$&L$\rm_{bol}$&$\lambda_{\rm Edd}$\\ 
    &deg&deg&erg~s$^{-1}$&erg~s$^{-1}$&\\
       \hline
        \hline
J0811&122.943967&2.526126&45.1&47.0&0.26\\
J1130&172.514848&36.677312&44.2&45.9&0.05\\
J0832&128.024828&52.733088&44.3&46.4&0.74\\
J1349&207.431154&26.019741&44.7&46.4&0.46\\
J1235&188.808355&12.457988&43.9&45.7&0.36\\
J1000&150.207954&2.083332&45.0&46.1&0.57\\
J1128&172.081646&58.628273&43.8&45.4&0.28\\
J1532&233.209209&30.287191&42.6&44.4&0.02\\
J1238&189.503915&62.226722&43.5&45.4&0.23\\
J1602&240.589793&43.054859&42.7&44.3&0.03\\
J1215&183.905837&47.231420&43.6&46.0&0.81\\
J1604&241.138799&43.060804&43.1&44.8&0.14\\
J1426&216.561798&35.104435&43.5&44.9&0.12\\
J1415&213.831276&-0.505989&43.2&44.4&0.21\\
       \hline
\end{tabular}
\end{center}
\end{table}
\endgroup

\begin{figure}
\centering
\includegraphics[width=.48\textwidth]{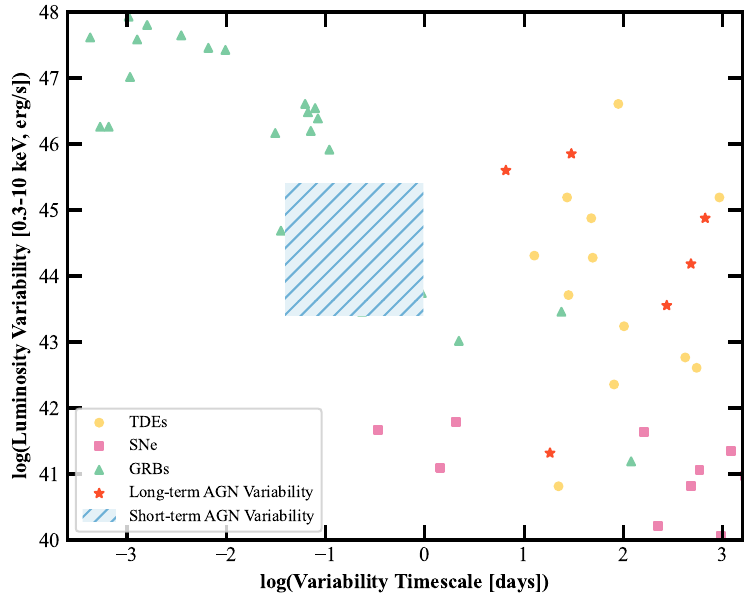}
\caption{The luminosity variability and variability timescale of X-ray transients and variables, adopted from \citet{Polzin2023}. We adopted rest frame time above half-maximum flux reported in \citet{Polzin2023} as the variability timescale. While sharing similar luminosity variability with TDEs and long-timescale AGN activity, the short-timescale variabilities of AGN coronae (blue hatched region) unfold on significantly shorter timescales, akin to those of some long GRBs. }\label{fig:transients}
\end{figure}

\section{A Rarely-Probed Class of X-ray Variable Phenomena}\label{sec:rare}
The characteristic timescales of the coronal variabilities identified in our sample are typically on the order of a few to tens of ks. During these episodes, the X-ray varied more than a factor of two, indicating variability luminosities comparable to the quiescent AGN X-ray luminosities, ranging from $\sim$10$^{43}$ to 10$^{45}$erg~s$^{-1}$. The X-ray luminosities were derived by fitting the {\it Chandra} spectra of each source. Spectra were extracted for each observation using the standard pipeline provided by {\tt CIAO} v4.15. The source and background extraction regions were identical to those used in the light curve analysis to ensure consistency.

Spectral modeling was performed using \texttt{XSPEC} v12.13.1 \citep{Arnaud1996}, adopting a standard model for unobscured AGN. This includes an absorbed power-law to represent the intrinsic coronal emission, potentially attenuated by line-of-sight dusty gas, and a blackbody component to account for the soft X-ray excess observed in some sources, though the physical origin of this component remains debated. Galactic absorption values were obtained using the \texttt{nh} tool \citep{nh}. It is worth noting that the derived X-ray luminosities represent time-averaged values across the full duration of each observation. The resulting 2--10~keV luminosities for each source are reported in Table~\ref{Table:source_property}. These events therefore represent a substantial release of energy on very short timescales, implying highly efficient dissipation mechanisms within the corona.

Comparable energy outputs have previously been observed in other extreme astrophysical transients, such as tidal disruption events (TDEs), long-term AGN variabilities, and some Gamma-ray bursts (GRBs). While displaying similar luminosity variability to TDEs and long-term AGN activity, processes that also involve massive energy dissipation around SMBHs, the short-term variabilities of AGN coronae unfold on significantly shorter timescales, comparable to some long GRBs.

In Fig.~\ref{fig:transients}, we place these AGN coronal variabilities in context by comparing their luminosity variabilities and characteristic variability timescale to those of other known classes of high-energy transients presented in \citet{Polzin2023}. This comparison highlights AGN coronal variabilities as a previously underexplored population of energetic, short-timescale events surrounding the SMBHs. 

\begingroup
\renewcommand*{\arraystretch}{1.2}
\begin{table}
\begin{center}
\caption{Global coronal size of a few AGN measured using microlensing methods. The black hole mass, half-light radius (HLR) of the corona, and coronal size (CS) of the sources are adopted from \citet{Chartas2016}.}\label{Table:CS_micro}
  \begin{tabular}{cccccccc}
       \hline
       \hline
    Source&$z$&log(M$\rm_{BH}$)&HLR&CS\\ 
    &&M$_{\odot}$&10$^{-4}$~pc&$r\rm _g$\\       
        \hline
RX J1131--1231&0.658&7.8$_{-0.6}^{+0.2}$&0.3$_{-0.1}^{+0.5}$&14$_{-6}^{+13}$\\   
Q J0158-4325&1.29&8.2$_{-0.1}^{+0.1}$&0.7$_{-0.5}^{+1.0}$&10$_{-5}^{+9}$\\    
SDSS 0924+0219&1.524&8.4$_{-0.5}^{+0.3}$&0.8$_{-0.6}^{+3.4}$&10$_{-6}^{+15}$\\
PG 1115+080&1.722&8.7$_{-0.3}^{+0.1}$&13$_{-11}^{+40}$&88$_{-50}^{+116}$\\            
HE 0435-1223&1.689&8.7$_{-0.4}^{+0.2}$&$<$5&$<$59\\    
HE 1104-1805&2.32&8.8$_{-0.3}^{+0.1}$&$<$7&$<$46\\  
Q 2237+0305&1.695&9.1$_{-0.4}^{+0.2}$&21$_{-13}^{+32}$&44$_{-19}^{+36}$\\
        \hline
         \hline     
\end{tabular}
\end{center}
\end{table}
\endgroup

\section{Bolometric Luminosity and Eddington Ratio}
The bolometric luminosity ($L_{\rm bol}$; integrated over 100~$\mu$m to 0.12~nm or 10~keV) and the Eddington ratio ($\lambda_{\rm Edd}$) are key indicators of the accretion state of SMBH. For SDSS AGN, $L_{\rm bol}$ values were adopted from \citet{Wu2022}, who estimated them by applying bolometric correction factors to continuum luminosities based on the mean AGN spectral energy distribution (SED) from \citet{Richards2006}. For DESI AGN, we derived bolometric luminosities using the same methodology and correction factors to ensure consistency. The Eddington ratio, defined as $\lambda_{\rm Edd} = L_{\rm bol} / (1.26 \times 10^{38}\,{\rm erg\,s^{-1}} , M_{\rm BH}/M_\odot)$, where $M_\odot$ is the Solar mass, quantifies the accretion rate relative to the Eddington limit. Both the bolometric luminosities and Eddington ratios for all sources are reported in Table~\ref{Table:source_property}.

\section{Coronal Size Derived using Microlensing Method}
The coronal size derived using the micro-lensing method \citep{Chartas2016} is reported in Table~\ref{Table:CS_micro}.

\section{Lightcurves of Variable AGN}
\begin{figure*}[h]
\centering
\includegraphics[width=.85\textwidth]{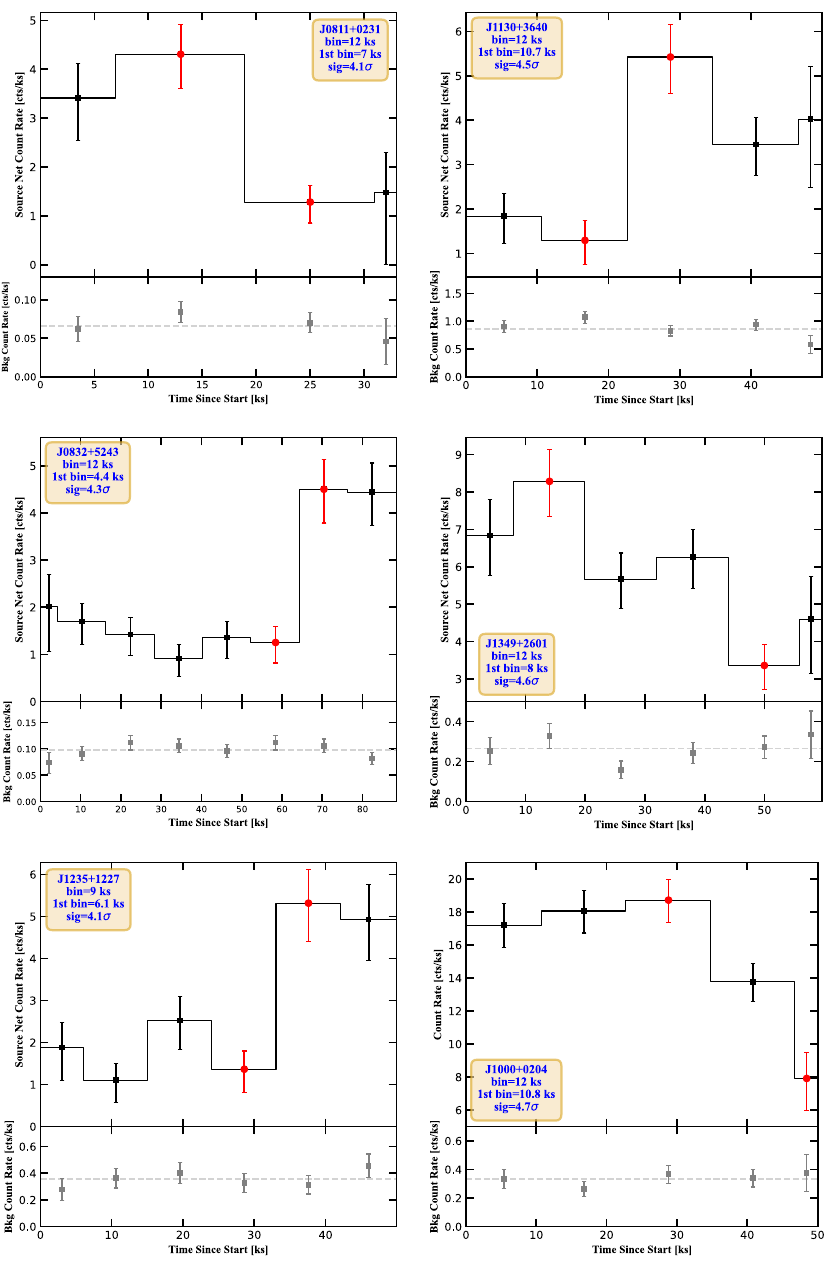}
\caption{Light curves in 0.5--7 keV of the 16 observations of the 14 sources studied in this work. The background count rate has been normalized to the source region size. The two bins used to compute t$_{\rm min}$ are plotted in red points.}
\label{fig:LC1}
\end{figure*}

\begin{figure*}[h]
\centering
\includegraphics[width=.85\textwidth]{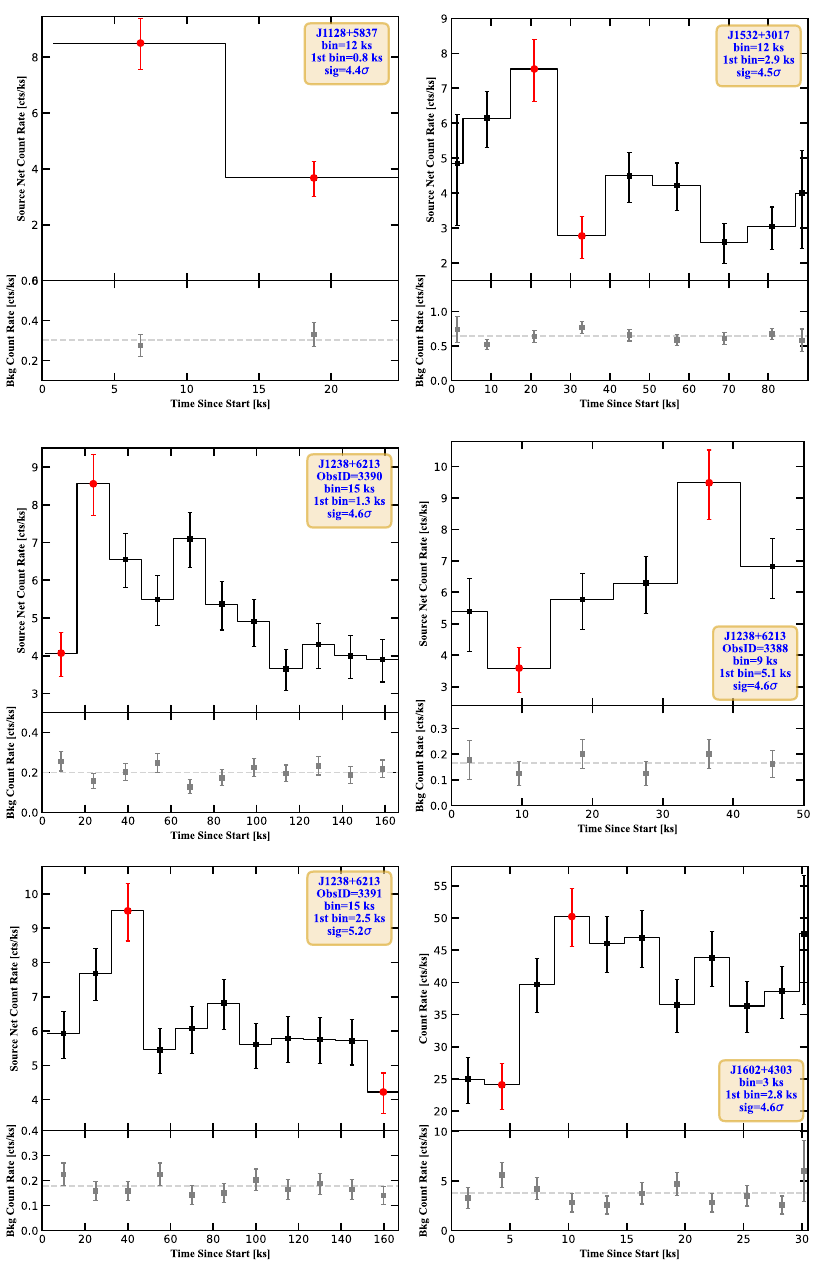}
\caption{Figure~\ref{fig:LC1} (continued). J1238+6213 presents significant variability in three separate observations.}
\label{fig:LC2}
\end{figure*}

\begin{figure*}[h]
\centering
\includegraphics[width=.85\textwidth]{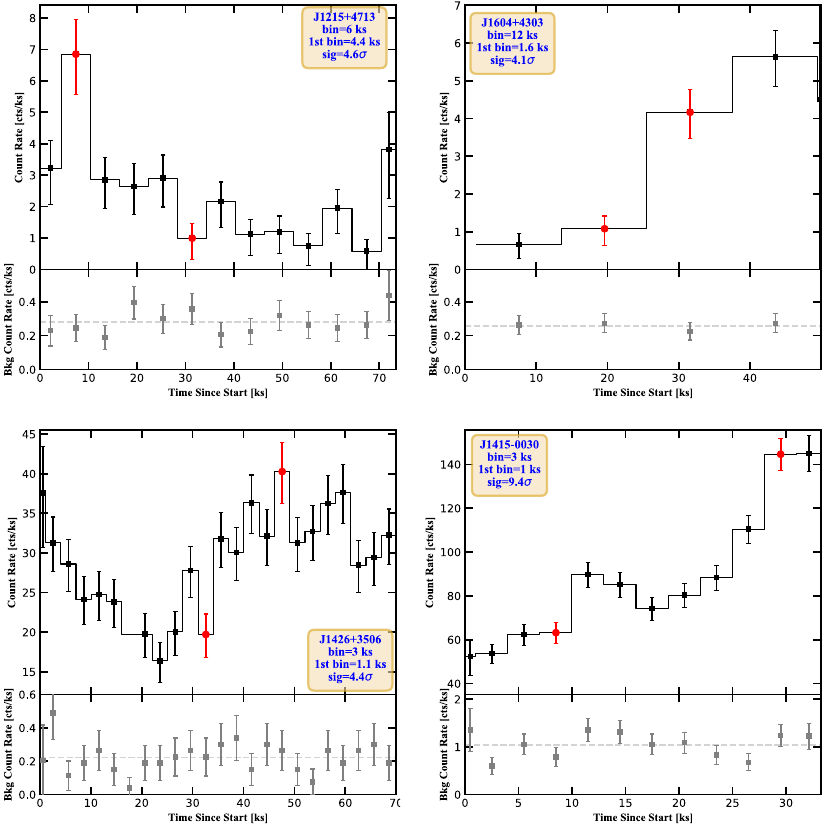}
\caption{Figure~\ref{fig:LC1} (continued).}
\label{fig:LC3}
\end{figure*}

The \cha\ lightcurves of the variable AGN selected in our sample are plotted in Fig.~\ref{fig:LC1}--\ref{fig:LC3}. The background count rate has been normalized to the source region size. The lightcurves are grouped according to the binning interval at which significant variability is detected. 

\bibliography{sn-bibliography}{}
\bibliographystyle{aasjournal}

\end{document}